\documentclass [aps,preprint,prd,superscriptaddress,showpacs,nofootinbib]{revtex4}
\usepackage{verbatim}
\usepackage{slashed}
\usepackage{epsfig}
\usepackage{axodraw}
\usepackage{graphicx}
\usepackage{amsmath}
\usepackage{mathrsfs}
\usepackage{bm}

\begin{document}
\title{Relativistic corrections to
$J/\psi$ exclusive and inclusive double charm production at B
factories}
\author{Zhi-Guo He}
\email{hzgzlh@gmail.com}
\author{Ying \surname{Fan}}
\email{fanying@pku.edu.cn} \affiliation{Department of Physics,
Peking University,
 Beijing 100871, China}
\author{Kuang-Ta Chao}
\email{ktchao@th.phy.pku.edu.cn} \affiliation{Department of Physics,
Peking University, Beijing 100871, China} \affiliation{China Center
of Advanced Science and Technology (World Laboratory), Beijing
100080, China}
\begin{abstract}
In order to clarify the puzzling problems in double charm
production, relativistic corrections at order $v^{2}$ to the
processes $e^{+}e^{-}\rightarrow J/\psi+\eta_{c}$ and
$e^{+}e^{-}\rightarrow J/\psi+c\overline{c}$ at B factories are
studied in non-relativistic quantum chromodynamics. The
short-distance parts of production cross sections are calculated
perturbatively, while the long-distance matrix elements are
estimated from $J/\psi$ and $\eta_c$ decays up to errors of order
$v^4$. Our results show that the relativistic correction to the
exclusive process $e^{+}e^{-}\rightarrow J/\psi+\eta_{c}$ is
significant, which, when combined together with the next-to-leading
order $\alpha_{s}$ corrections, could resolve the large discrepancy
between theory and experiment; whereas for the inclusive process
$e^{+}e^{-}\rightarrow J/\psi+c\overline{c}$ the relativistic
correction is tiny and negligible. The physical reason for the above
difference between exclusive and inclusive processes largely lies in
the fact that in the exclusive process the relative momentum between
quarks in charmonium substantially reduces the virtuality of the
gluon that converts into a charm quark pair, but this is not the
case for the inclusive process, in which the charm quark
fragmentation $c\to J/\psi+c$ is significant, and QCD radiative
corrections can be more essential.

\end{abstract}


\maketitle

\section{Introduction}

The exclusive and inclusive double charm production in $e^{+} e^{-}$
annihilation at $B$ factories has been one of the most puzzling
problems in nonrelativistic QCD (NRQCD)~\cite{9407339} and heavy
quarkonium physics~\cite{0412158} for years. The measured cross
sections for exclusive double charmonium production $e^{+} e^{-}\to
J/\psi +\eta_c$ and inclusive double charm production $e^{+}
e^{-}\to J/\psi +c\bar c$ at $\sqrt{s}=10.6$ GeV are much higher
than the leading order (LO) predictions in NRQCD.

The cross section of $e^{+} e^{-}\to J/\psi +\eta_c$ at
$\sqrt{s}=10.6$ GeV observed by Belle\cite{Bell1,0412041} is

\begin{subequations}\label{eq1}
\begin{equation} \label{eqa}
\sigma[e^{+}e^{-}\rightarrow J/\psi+\eta_{c}]\times
\mathcal{B}^{\eta_{c}}[\geq2]=25.6\pm2.8\pm3.4~fb,
\end{equation}
and by BaBar\cite{BARBAR} is
\begin{equation}\label{eqb}
\sigma[e^{+}e^{-}\rightarrow J/\psi+\eta_{c}]\times
\mathcal{B}^{\eta_{c}}[\geq2]=17.6\pm2.8\pm2.1~fb,
\end{equation}
\end{subequations}
where $\mathcal{B}^{\eta_{c}}[\geq2]$ is the branching fraction for
$\eta_{c}$ decay into at least two charged particles. Since
$\mathcal{B}^{\eta_{c}}[\geq2]<1$, Eq.[\ref{eq1}] is the lower bound
of the cross section of the double charmonium production. It is
about an order of magnitude larger than the theoretical
predictions\cite{0211085,0211181,0305102}, at the leading order of
strong coupling constant $\alpha_s$ and quark relative velocity $v$
in the NRQCD factorization approach\cite{9407339}. In\cite{0506076}
the next-to leading order (NLO) QCD result is given and it is found
that the $\mathcal {O}(\alpha_s)$ corrections may greatly reduce the
large discrepancy between theory and experiment. Meanwhile, the
relativistic corrections are also considered by a number of authors.
In \cite{0605230}, relativistic and bound state effects are
discussed on the basis of relativistic quark models. In
\cite{0405111,0412335,0506009,0701234,0702147}, relativistic effects
are estimated in the framework of the light cone method, and very
large enhancement effects on the cross sections can be found with
certain light-cone distribution amplitudes. In \cite{0603186}, the
authors calculate the relativistic correction based on NRQCD and
light cone method with long-distance matrix elements determined from
the Cornell potential model, and they point out, however, that after
subtracting parts of the light-cone distribution functions that
correspond to corrections of relative-order $\alpha_s$ in NRQCD, the
enhancement effect due to relativistic corrections is not very large
but still substantial. Probably,
with both relative-order-$\alpha_s$  and relativistic corrections we
may resolve the discrepancy between theory and experiment for
$e^{+}e^{-}\rightarrow J/\psi+\eta_{c}$\cite{0603186} .

The large discrepancies between theory and experiment exist not only
in the exclusive double charmonium production processes such as
$e^{+}e^{-}\rightarrow J/\psi+\eta_{c}$ but also in the inclusive
production process of $e^{+}e^{-}\rightarrow J/\psi+c\bar{c}$. The
experimental result measured by Belle\cite{Bell1} is
\begin{equation} \label{eq2}
\sigma[e^{+}e^{-}\rightarrow
J/\psi+c\bar{c}+X]=0.87^{+0.21}_{-0.19}\pm0.17\textrm{pb},
\end{equation}
which is about five times larger than the theoretical
calculations\cite{0301218,9606229,Yuan,Lee,rusia2} based on NRQCD at
leading order of $\alpha_{s}$ and $v^{2}$. In~\cite{0305084} the
two-photon contribution  and in~\cite{0301218} the color-octet
contribution to $J/\psi+c\bar{c}$ production are further considered,
but they are too small to resolve the large discrepancy. Other
attempts to solve the problem can be found in a comprehensive review
on heavy quarkonium physics~\cite{0412158}. It is certainly
interesting to see whether the large discrepancy can be resolved by
inclusion of NLO QCD corrections and relativistic corrections.
Recently, in ~\cite{0611086}, the authors calculate the prompt
$J/\psi+c\bar{c} +X$ production at NLO $\alpha_{s}$ including direct
production and feeddown contribution mainly from the $\psi(2S)$, and
find the NLO $\alpha_{s}$ corrections are large and positive, and
could be helpful to settle the problem between experiment and
theory. But we still need to know how large are the relativistic
corrections, and whether they are positive or negative to the
solution of the problem.

In this paper, we consider the relativistic corrections to both
these exclusive and inclusive double-charm production processes
based on NRQCD formulas in the color-singlet sector, since the
color-octet contributions are negligible~\cite{0301218}. In order to
avoid the model dependence in determining the long distance matrix
elements, differing from \cite{0603186}, we determine the matrix
elements of up to dimension-8 four fermion operators from the
observed decay rates of $J/\psi$ and $\eta_c$. We find that the
relativistic effect on the double charmonium production
$e^{+}e^{-}\rightarrow J/\psi+\eta_{c}$ is substantial and
comparable to the estimate of \cite{0603186}; whereas for the
inclusive production $e^{+}e^{-}\rightarrow J/\psi+c\bar{c}$ the
relativistic corrections are very small and negligible.  The rest of
this paper is organized as follows. In Section II, we give the
general formulas of the production rates in NRQCD at $v^{2}$ order.
Relativistic corrections to the exclusive process and the inclusive
process are studied in Section III and Section IV respectively. A
summary for the $v^{2}$ order corrections to these two processes
will be given in Section V. In the Appendix we give some of the
analytic results.

\section{Production Cross Sections in NRQCD}

In NRQCD the production and decay of charmonia are factorized into
two parts, the short distance part that can be calculated
perturbatively, and the long distance part can be estimated by
lattice calculation, phenomenological models, or from other
experimental observables. The long distance parts are related to the
four fermion operators, characterized by the velocity $v$ of the
charm quark  in the meson rest frame. The production cross sections
of $\eta_{c}$ and $J/\psi$ up to $v^{2}$ order are \cite{9407339}
\begin{subequations}
\begin{equation}
\sigma(\eta_{c})=\frac{F_{1}({}^{1}S_{0})}{m_{c}^2}\langle
0|\mathcal{O}_{1}({}^{1}S_{0}^{\eta_{c}})|0\rangle+\frac{G_{1}({}^{1}S_{0})}{m_{c}^4}\langle
0|\mathcal{P}_{1}({}^{1}S_{0}^{\eta_{c}})|0\rangle+O(v^{4}\sigma),
\end{equation}
\begin{equation}
\sigma(\psi)=\frac{F_{1}({}^{3}S_{1})}{m_{c}^2}\langle
0|\mathcal{O}_{1}({}^{3}S_{1}^{\psi})|0\rangle+\frac{G_{1}({}^{3}S_{1})}{m_{c}^4}\langle
0|\mathcal{P}_{1}({}^{3}S_{1}^{\psi})|0\rangle+O(v^{4}\sigma).
\end{equation}
\end{subequations}
  The operators are defined as
\begin{subequations}
\begin{equation}
\mathcal{O}_{1}^{\eta_{c}}({}^{1}S_{0})=
\chi^{\dagger}\psi(a_{\eta_{c}}^{\dagger}a_{\eta_{c}})\psi^{\dagger}\chi,
\end{equation}
\begin{equation}
\mathcal{P}_{1}^{\eta_{c}}({}^{1}S_{0})=\frac{1}{2}[\chi^{\dagger}\psi(a_{\eta_{c}}^{\dagger}a_{\eta_{c}})
\psi^{\dagger}(-\frac{i}{2}\overleftrightarrow{\mathbf{D}})^2\chi+
\chi^{\dagger}(-\frac{i}{2}\overleftrightarrow{\mathbf{D}})^2\psi(a_{\eta_{c}}^{\dagger}a_{\eta_{c}})
\psi^{\dagger}\chi],
\end{equation}
\begin{equation}
\mathcal{O}_{1}^{\psi}({}^{3}S_{1})=\chi^{\dagger}\sigma^{i}
\psi(a_{\psi}^{\dagger}a_{\psi})\psi^{\dagger}\sigma^{i}\chi,
\end{equation}
\begin{equation}
\mathcal{P}_{1}^{\psi}({}^{3}S_{1})=\frac{1}{2}[\chi^{\dagger}\sigma^{i}\psi(a_{\psi}^{\dagger}a_{\psi})
\psi^{\dagger}\sigma^{i}(-\frac{i}{2}\overleftrightarrow{\mathbf{D}})^2\chi+
\chi^{\dagger}\sigma^{i}(-\frac{i}{2}\overleftrightarrow{\mathbf{D}})^2\psi(a_{\psi}^{\dagger}a_{\psi})
\psi^{\dagger}\sigma^{i}\chi].
\end{equation}
\end{subequations}
  The short distance coefficients can be evaluated by the matching
condition:
\begin{equation}\label{mat}
\sigma(Q\overline{Q})\Big{|}_{\textrm{pert
QCD}}=\sum_{n}\frac{F_{n}(\Lambda)}{M^{d_{n}-4}}\langle0|\mathcal{O}_{n}^{Q\overline{Q}}
(\Lambda)|0\rangle\Big{|}_{\textrm{pert NRQCD}}
\end{equation}
  The left hand side of Eq. [\ref{mat}] can be calculated by the spinor
projection method \cite{Ku}. Furthermore (see,
e.g.,\cite{0205210}\cite{Jiang}), the projection of
$v(P/2-q)\overline{u}(P/2+q)$ onto a particular angular momentum
state can be expressed in a Lorentz covariant form. The momenta of
quark and antiquark in an arbitrary frame are
respectively~\cite{9604237}:
\begin{subequations}
\begin{equation}
\frac{1}{2}P+q=L(\frac{1}{2}P_{r}+\mathbf{q}),
\end{equation}
\begin{equation}
\frac{1}{2}P-q=L(\frac{1}{2}P_{r}-\mathbf{q}),
\end{equation}
\end{subequations}
where $P_{r}^{\mu}=(2E_{q},\mathbf{0})$,
$E_{q}=\sqrt{m^{2}+\mathbf{q}^{2}}$, and $2\mathbf{q}$ is the
relative momentum between two quarks in the meson rest frame.
$L_{\mu}^{v}$ is the boost tensor from the meson rest frame to an
arbitrary frame.

In the meson rest frame the expression of projection onto a state of
$S=0$ is
\begin{eqnarray}
\sum_{\lambda_{1}\lambda_{2}}v(\mathbf{-q},\lambda_{2})\overline{u}(\mathbf{q},\lambda_{1})\langle\frac{1}{2},\lambda_{1},
\frac{1}{2},\lambda_{2}|0,0\rangle=
\nonumber \\
\frac{1}{\sqrt{2}}(E+m)(1-\frac{\mbox{\boldmath$\alpha$}\cdot\mathbf{q}}{E+m})
\gamma_{5}\frac{1+\gamma_{0}}{2}(1+\frac{\mbox{\boldmath$\alpha$}\cdot\mathbf{q}}{E+m})\gamma_{0}.
\end{eqnarray}
In an arbitrary frame it becomes
\begin{eqnarray} \label{eq7}
\sum_{\lambda_{1}\lambda_{2}}v(q,\lambda_{2})\overline{u}(q,\lambda_{1})\langle\frac{1}{2},\lambda_{1},
\frac{1}{2},\lambda_{2}|0,0\rangle=
 \nonumber \\
-\frac{1}{2\sqrt{2}(E+m)}(\frac{1}{2}\slashed{P}-\slashed{q}-m)\gamma_{5}\frac{\slashed{P}+2E}{2E}
(\frac{1}{2} \slashed{P}+\slashed{q}+m).
\end{eqnarray}

Similarly, expressions of projection onto a state of $S=1$ in the
rest frame of the meson and an arbitrary frame are:
\begin{subequations}
\begin{eqnarray}\label{eq8a}
\sum_{\lambda_{1}\lambda_{2}}v(\mathbf{-q},\lambda_{2})\overline{u}(\mathbf{q},\lambda_{1})\langle\frac{1}{2},\lambda_{1};
\frac{1}{2},\lambda_{2}|1,\epsilon\rangle=
 \nonumber \\
\frac{1}{\sqrt{2}}(E+m)(1-\frac{\mbox{\boldmath$\alpha$}\cdot\mathbf{q}}{E+m})
{\mbox{\boldmath$\alpha$}}\cdot{\mbox{\boldmath$\epsilon$}}\frac{1+\gamma_{0}}{2}(1+\frac{\mbox{\boldmath$\alpha$}\cdot\mathbf{q}}{E+m})\gamma_{0},
\end{eqnarray}
\begin{eqnarray}\label{eq8b}
\sum_{\lambda_{1}\lambda_{2}}v(q,\lambda_{2})\overline{u}(q,\lambda_{1})\langle\frac{1}{2},\lambda_{1};
\frac{1}{2},\lambda_{2}|1,\epsilon\rangle=
 \nonumber \\
-\frac{1}{2\sqrt{2}(E+m)}(\frac{1}{2}\slashed{P}-\slashed{q}-m)\slashed{\epsilon}\frac{\slashed{P}+2E}{2E}
(\frac{1}{2} \slashed{P}+\slashed{q}+m).
\end{eqnarray}
\end{subequations}
In our calculation the Dirac spinors are normalized as
$\overline{u}u=-\overline{v}v=2m_{c}$. Then the short distance part
of the cross section can be calculated at any order of $v$.

At order $v^{2}$ the cross section of $e^{+}e^{-}\rightarrow
J/\psi+\eta_{c}$ and $e^{+}e^{-}\rightarrow J/\psi+c\overline{c}$
can be expressed as
\begin{subequations}
\begin{eqnarray}
\sigma(e^{+}e^{-}\rightarrow J/\psi+\eta_{c})=&\langle
0|\mathcal{O}_{1}({}^{1}S_{0}^{\eta_{c}})|0\rangle\frac{\langle
0|\mathcal{O}_{1}({}^{3}S_{1}^{\psi})|0\rangle}{3}
\frac{1}{2s}\int\overline{M_{0}}\;d \;LIPS+\nonumber\\
&\frac{\langle0|\mathcal{P}_{1}({}^{3}S_{1}^{\psi})|0\rangle}{3}
\langle0|\mathcal{O}_{1}({}^{1}S_{0}^{\eta_{c}})|0\rangle\frac{1}{2s}\int\overline{M}_{1J/\psi}\;d
\;LIPS+\nonumber\\
&\langle0|\mathcal{P}_{1}({}^{1}S_{0}^{\eta_{c}})|0\rangle\frac{\langle
0|\mathcal{O}_{1}({}^{3}S_{1}^{\psi})|0\rangle}{3}
\frac{1}{2s}\int\overline{M}_{1\eta_{c}}\;d \;LIPS \;.
\end{eqnarray}
\begin{align}
&\sigma(e^{+}e^{-}\rightarrow J/\psi+c\overline{c})=
\nonumber\\
&\frac{\langle0|\mathcal{O}_{1}({}^{3}S_{1}^{\psi})|0\rangle}{3}
\frac{1}{2s}\int\overline{N_{0}}\;d \;LIPS+
\frac{\langle0|\mathcal{P}_{1}({}^{3}S_{1}^{\psi})|0\rangle}{3}
\frac{1}{2s}\int\overline{N}_{1}\;d \;LIPS \;.
\end{align}
\end{subequations}
Here $LIPS$ means the Lorentz invariant phase space, bar means
averaging spins over the initial states and summing spins over the
final states, and
$\overline{M_{0}},\overline{M}_{1J/\psi},\overline{M}_{1\eta_{c}},\overline{N_{0}},\overline{N_{1}}$
can be calculated perturbatively.

\section{Relativistic Corrections to $e^{+}e^{-}\rightarrow J/\psi+\eta_{c}$}

\subsection{Short distance part}

  There are four Feynman diagrams in the process
$e^{+}e^{-}\rightarrow
(c\overline{c})_{{}^{3}S_{1}}+(c\overline{c})_{{}^{1}S_{0}}$, shown
in Fig.[1].
\begin{figure}[t]
\begin{center}
\includegraphics[width=11.5cm]{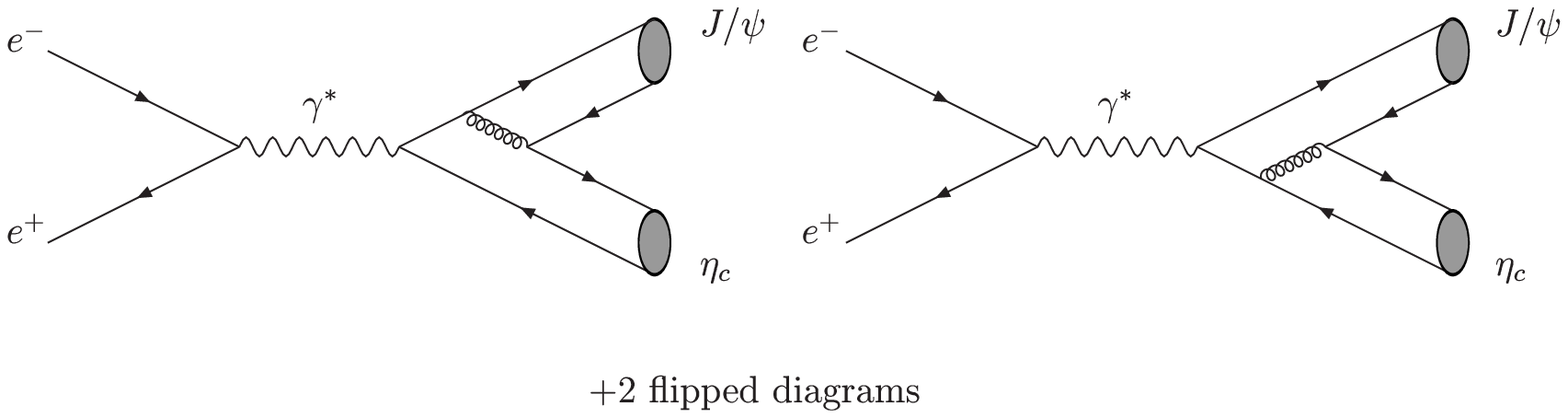}
\end{center}
\caption{$e^{+}e^{-}\rightarrow
(c\overline{c})_{{}^{3}S_{1}}+(c\overline{c})_{{}^{1}S_{0}}$}
\end{figure}
The amplitude of the process can be expanded in terms of the quark
relative momentum in charmonium:
\begin{eqnarray}
M(e^{+}e^{-}\rightarrow
(c\overline{c})_{{}^{3}S_{1}}+(c\overline{c})_{{}^{1}S_{0}})=
(\frac{m_{c}}{E_{1}}\frac{m_{c}}{E_{2}})^{1/2}A(q_{\psi},q_{\eta_{c}})=\nonumber\\
(\frac{m_{c}}{E_{1}}\frac{m_{c}}{E_{2}})^{1/2}
(A(0,0)+q_{\psi}^{\alpha}\frac{\partial A}{\partial
q_{\psi}^{\alpha}}\Big{|}_{q_{\psi}=q_{\eta_{c}}=0}
+q_{\eta_{c}}^{\alpha}\frac{\partial A}{\partial
q_{\eta_{c}}^{\alpha}}\Big{|}_{q_{\psi}=q_{\eta_{c}}=0}+ \nonumber\\
\frac{1}{2}q_{\psi}^{\alpha}q_{\psi}^{\beta}\frac{\partial^{2}
A}{\partial q_{\psi}^{\alpha}\partial
q_{\psi}^{\beta}}\Big{|}_{q_{\psi}=q_{\eta_{c}}=0}+
\frac{1}{2}q_{\eta_{c}}^{\alpha}q_{\eta_{c}}^{\beta}\frac{\partial^{2}
A}{\partial q_{\eta_{c}}^{\alpha}\partial
q_{\eta_{c}}^{\beta}}\Big{|}_{q_{\psi}=q_{\eta_{c}}=0}+\dots),
\end{eqnarray}
and $A(q_{\psi},q_{\eta_{c}})$ is expressed as
\begin{eqnarray}
A(q_{\psi},q_{\eta_{c}})=\sum_{\lambda_{1}\lambda_{2}\lambda_{3}\lambda_{4}}\sum_{ijkl}
\langle\frac{1}{2},\lambda_{1};
\frac{1}{2},\lambda_{2}|1,S_{z}\rangle
\langle\frac{1}{2},\lambda_{3}; \frac{1}{2},\lambda_{4}|0,0\rangle
\langle3,i;\overline{3},j|1\rangle\langle3,k;\overline{3},l|1\rangle\nonumber\\
A(e^{+}e^{-}\rightarrow
c_{\lambda1,i}(\frac{P1}{2}+q_{\psi})\overline{c}_{\lambda2,j}(\frac{P1}{2}-q_{\psi})+
c_{\lambda3,k}(\frac{P2}{2}+q_{\eta_{c}})\overline{c}_{\lambda4,l}(\frac{P2}{2}-q_{\eta_{c}})),
\nonumber\\
\end{eqnarray}
where $\langle 3,i;\bar{3},j|1\rangle =\delta_{ij}/\sqrt{N_c}$ and
$\langle 3,k;\bar{3},l|1\rangle=\delta_{kl}/\sqrt{N_c}$ are the
color-SU(3), Clebsch-Gordon coefficients for $Q\bar{Q}$ pairs
projecting onto a color singlet state. Using Eq.[\ref{eq7}] and
Eq.[\ref{eq8b}], we can express $A(q_{\psi},q_{\eta_{c}})$ in a
covariant form. The factor
$\displaystyle(\frac{m_{c}}{E_{1}}\frac{m_{c}}{E_{2}})^{1/2}$ comes
from the relativistic normalization of the $c\overline{c}$ state,
and
$E_{1}=\sqrt{m_{c}^2+\mathbf{q_{\psi}}^2},E_{2}=\sqrt{m_{c}^2+\mathbf{q_{\eta_{c}}}^2}$.

For the $S$ wave charmonium
$q^{\alpha}q^{\beta}=\frac{1}{3}\mathbf{q}^{2}(-g^{\alpha
\beta}+\frac{P^{\alpha}P^{\beta}}{P^{2}})$, where $P^2=4E^2, P\cdot
q=0$ and the odd-power terms of $v$ vanish. Then at leading order of
$v^{2}$
\begin{align}
|M|^{2}=&\frac{m_{c}}{E_{1}}\frac{m_{c}}{E_{2}}A(0,0)A^{\ast}(0,0)+
\frac{1}{2}q_{\psi}^{\alpha}q_{\psi}^{\beta}A_{\psi}^{\alpha\beta}A^{\ast}(0,0)
+\frac{1}{2}q_{\eta_{c}}^{\alpha}q_{\eta_{c}}^{\beta}A_{\eta_{c}}^{\alpha\beta}A^{\ast}(0,0)+
\nonumber\\
&\frac{1}{2}q_{\psi}^{\alpha}q_{\psi}^{\beta}A_{\psi}^{\ast\alpha\beta}A(0,0)
+\frac{1}{2}q_{\eta_{c}}^{\alpha}q_{\eta_{c}}^{\beta}A_{\eta_{c}}^{\ast\alpha\beta}A(0,0),
\end{align}
where $A^{\alpha\beta}=\frac{\partial^{2}A}{\partial
q^{\alpha}\partial q^{\beta}}$,
$A^{\ast\alpha\beta}=\frac{\partial^{2}A^{\ast}}{\partial
q^{\alpha}\partial q^{\beta}}$.

To obtain the short distance part
$E=\sqrt{m_{c}^{2}+\mathbf{q}^{2}}$ should also be expanded in
$v^{2}=\frac{\mathbf{q}^{2}}{m_{c}^{2}}$. Then the relation of
$M_{0}, M_{1J/\psi}, M_{1\eta_{c}}$ to $|M|^{2}$ can be easily
written down. Details of how to get the short distance coefficients
from the covariant projection method can be found in e.g.
\cite{9602223}
\begin{subequations}
\begin{equation}
\overline{M}_{0}=\frac{1}{(2N_{c}m_{c})^{2}}(A(0,0)A^{\ast}(0,0)
)\Big{|}_{\mathbf{q}_{J/\psi}^{2}=\mathbf{q}_{\eta_{c}}^{2}=0};
\end{equation}
\begin{align}
\overline{M}_{1J/\psi}=&\frac{1}{(2N_{c}m_{c})^{2}}(\frac{\partial
(\frac{m_{c}^{2}}{E_{1}E_{2}}A(0,0)A^{\ast}(0,0))}{\partial
(\mathbf{q}_{\psi}^{2})}\Big{|}_{\mathbf{q}_{J/\psi}^{2}=\mathbf{q}_{\eta_{c}}^{2}=0}
\nonumber\\+&\frac{1}{6}(A_{\psi}^{\alpha\beta}\Pi_{\alpha\beta}(P_{J/\psi})A^{\ast}(0,0)+
A_{\psi}^{\ast\alpha\beta}\Pi_{\alpha\beta}(P_{J/\psi})A(0,0))
)\Big{|}_{\mathbf{q}_{J/\psi}^{2}=\mathbf{q}_{\eta_{c}}^{2}=0};
\end{align}
\begin{align}
\overline{M}_{1\eta_{c}}=&\frac{1}{(2N_{c}m_{c})^{2}}(\frac{\partial
(\frac{m_{c}^{2}}{E_{1}E_{2}}A(0,0)A^{\ast}(0,0))}{\partial
(\mathbf{q}_{\eta_{c}}^{2})}\Big{|}_{\mathbf{q}_{J/\psi}^{2}=\mathbf{q}_{\eta_{c}}^{2}=0}
\nonumber\\+&\frac{1}{6}(A_{\eta_{c}}^{\alpha\beta}\Pi_{\alpha\beta}(P_{\eta_{c}})A^{\ast}(0,0)+
A_{\eta_{c}}^{\ast\alpha\beta}\Pi_{\alpha\beta}(P_{\eta_{c}})A(0,0)))
\Big{|}_{\mathbf{q}_{J/\psi}^{2}=\mathbf{q}_{\eta_{c}}^{2}=0}.
\end{align}
\end{subequations}
Here $\overline{M}_{0}$ is just the leading order result, and it
agrees with the previous result in ref.\cite{0211181}
\begin{equation}
\overline{M}_{0}=\frac{1}{9N_{c}^2}
\frac{2048(s-16m_{c}^{2})(1+\cos^{2}(\theta))(4\pi\alpha)^{2}(4\pi\alpha_{s})^{2}e_{q}^{2}}{s^4}.
\end{equation}
The expressions of $\overline{M}_{1J/\psi}$ and
$\overline{M}_{1\eta_{c}}$ are
\begin{subequations}
\begin{align}
\overline{M}_{1J/\psi}&=\frac{(1+\cos^{2}_\theta)(4\pi\alpha)^{2}(4\pi\alpha_{s})^{2}e_{q}^{2}}{9N_{c}^{2}}
(\frac{512(16m_{c}^{2}-3s)}{m_{c}^{2}s^{4}}+
\frac{512(2560m_{c}^{4}-592m_{c}^{2}s+27s^2)}{3m_{c}^{2}s^{5}}),
\end{align}
\begin{eqnarray}
\overline{M}_{1\eta_{c}}=\frac{(1+\cos^{2}_\theta)(4\pi\alpha)^{2}(4\pi\alpha_{s})^{2}e_{q}^{2}}{9N_{c}^2}
(\frac{-16384}{s^{4}}+\frac{1024(11s-80m_{c}^{2})(s-16m_{c}^{2})}{3m_{c}^{2}s^{5}}),
\end{eqnarray}
\end{subequations}
where $Nc=3$, $e_{q}=\frac{2}{3}$, and $\theta$ is the angle between
the $J/\psi$ and the electron.

\subsection{Long distance part}
In this section we present a phenomenological estimate of the
color-singlet production matrix elements which are extracted from
charmonium decay data. It is known that up to errors of order $v^4$
the color-singlet production matrix elements are related to the
decay matrix elements through vacuum saturation~\cite{9407339}.
There are three independent NRQCD matrix elements at order $v^{2}$,
$\langle0|\mathcal{O}_{1}({}^{1}S_{0}^{\eta_{c}})|0\rangle$,
$\langle0|\mathcal{O}_{1}({}^{3}S_{1}^{\psi})|0\rangle$, and
$\langle0|\mathcal{P}_{1}({}^{1}S_{0}^{\eta_{c}})|0\rangle=
\langle0|\mathcal{P}_{1}({}^{3}S_{1}^{\psi})|0\rangle/3$
$(1+O(v^{2}))$. We can get them through the $J/\psi$ leptonic decay
$J/\psi \rightarrow e^{+}e^{-}$ and hadronic decay $J/\psi
\rightarrow LH$, and the $\eta_{c}$ photonic decay
$\eta_{c}\rightarrow \gamma\gamma$. The theoretical results at
next-to-leading order of $\alpha_{s}$ and $v^{2}$ for $J/\psi
\rightarrow e^{+}e^{-}$, $J/\psi \rightarrow  LH$ \footnote{We do
not include the electromagnetic process of $J/\psi\rightarrow
\gamma^{*}\rightarrow LH.$ }, and $\eta_{c}\rightarrow \gamma\gamma$
are summarized in ref.\cite{0205210} as
\begin{subequations}
\begin{equation}
\Gamma[\eta_{c}\rightarrow \gamma\gamma]=2e_{c}^{4}\pi\alpha^{2}
\Big((1-\frac{(20-\pi^{2})\alpha_{s}}{3\pi})\frac{\langle0|\mathcal{O}_{1}({}^{1}S_{0}^{\eta_{c}})|0\rangle}{m_{c}^{2}}
-\frac{4}{3}\frac{\langle0|\mathcal{P}_{1}({}^{1}S_{0}^{\eta_{c}})|0\rangle}{m_{c}^{4}}\Big),
\end{equation}
\begin{equation}
\Gamma[J/\psi\rightarrow
e^{+}e^{-}]=\frac{2e_{c}^{2}\pi\alpha^{2}}{3}
\Big((1-\frac{16\alpha_{s}}{3\pi})\frac{\langle0|\mathcal{O}_{1}({}^{3}S_{1}^{\psi})|0\rangle/3}{m_{c}^{2}}-
\frac{4}{3}\frac{\langle0|\mathcal{P}_{1}({}^{3}S_{1}^{\psi})|0\rangle/3}{m_{c}^{4}}\Big),
\end{equation}
\begin{equation}
\Gamma[J/\psi\rightarrow
LH]=(\frac{20\alpha_{s}^{3}}{243}(\pi^{2}-9))
\Big((1-2.55\frac{\alpha_{s}}{\pi})\frac{\langle0|\mathcal{O}_{1}({}^{3}S_{1}^{\psi})|0\rangle/3}{m_{c}^{2}}-
\frac{19\pi^{2}-132}{12\pi^{2}-108}
\frac{\langle0|\mathcal{P}_{1}({}^{3}S_{1}^{\psi})|0\rangle/3}{m_{c}^{4}}\Big).
\end{equation}
\end{subequations}
Solving these equations at leading order of $\alpha_{s}$ (QCD
radiative corrections not included), we get
\begin{subequations}\label{eqM1}
\begin{equation}
\langle0|\mathcal{O}_{1}({}^{1}S_{0}^{\eta_{c}})|0\rangle=0.286GeV^{3},
\langle0|\mathcal{O}_{1}({}^{3}S_{1}^{\psi})|0\rangle/3=0.295GeV^{3},
\end{equation}
\begin{equation}
\frac{\langle0|\mathcal{P}_{1}({}^{1}S_{0}^{\eta_{c}})|0\rangle}{m_{c}^{2}}=
\frac{\langle0|\mathcal{P}_{1}({}^{3}S_{1}^{\psi})|0\rangle}{3m_{c}^{2}}=0.321
\times10^{-1}GeV^{3},
\end{equation}
\end{subequations}
for $m_{c}=$1.5Gev and $\alpha_{s}=0.26$. The experimental data of
these decay rates can be found from \cite{PDG}, and we choose their
central values $\Gamma[J/\psi\rightarrow e^{+}e^{-}]=5.55$KeV,
$\Gamma[J/\psi\rightarrow LH]=69.3$KeV, and
$\Gamma[\eta_{c}\rightarrow\gamma\gamma]=7.14$KeV. The matrix
elements can be expressed as functions of the charm quark mass,
\begin{subequations}\label{eqM2}
\begin{eqnarray}
\langle0|\mathcal{O}_{1}({}^{1}S_{0}^{\eta_{c}})|0\rangle=0.127m_{c}^{2},
\end{eqnarray}
\begin{eqnarray}
\frac{\langle0|\mathcal{O}_{1}({}^{3}S_{1}^{\psi})|0\rangle}{3}=0.131m_{c}^{2},
\end{eqnarray}
\begin{eqnarray}
\frac{\langle0|\mathcal{P}_{1}({}^{1}S_{0}^{\eta_{c}})|0\rangle}{m_{c}^{2}}=
\frac{\langle0|\mathcal{P}_{1}({}^{3}S_{1}^{\psi})|0\rangle}{3m_{c}^{2}}=0.014m_{c}^{2}
\end{eqnarray}
\end{subequations}
Including the QCD NLO radiative corrections, and doing the
calculation in the same way as above, we have
\begin{subequations}\label{eqM3}
\begin{equation}
\langle0|\mathcal{O}_{1}({}^{1}S_{0}^{\eta_{c}})|0\rangle=0.432GeV^{3},
\langle0|\mathcal{O}_{1}({}^{3}S_{1}^{\psi})|0\rangle/3=0.573GeV^{3},
\end{equation}
\begin{equation}
\frac{\langle0|\mathcal{P}_{1}({}^{1}S_{0}^{\eta_{c}})|0\rangle}{m_{c}^{2}}=
\frac{\langle0|\mathcal{P}_{1}({}^{3}S_{1}^{\psi})|0\rangle}{3m_{c}^{2}}=0.514
\times10^{-1}GeV^{3},
\end{equation}
\end{subequations}
for $m_{c}=$1.5Gev and $\alpha_{s}=0.26$. And we then have
\begin{subequations}\label{eqM4}
\begin{eqnarray}
\langle0|\mathcal{O}_{1}({}^{1}S_{0}^{\eta_{c}})|0\rangle=0.192m_{c}^{2},
\end{eqnarray}
\begin{eqnarray}
\frac{\langle0|\mathcal{O}_{1}({}^{3}S_{1}^{\psi})|0\rangle}{3}=0.255m_{c}^{2},
\end{eqnarray}
\begin{eqnarray}
\frac{\langle0|\mathcal{P}_{1}({}^{1}S_{0}^{\eta_{c}})|0\rangle}{m_{c}^{2}}=
\frac{\langle0|\mathcal{P}_{1}({}^{3}S_{1}^{\psi})|0\rangle}{3m_{c}^{2}}=0.023m_{c}^{2},
\end{eqnarray}
\end{subequations}
as functions of $m_{c}$.

In \cite{9701353}, the authors express the next-to-leading order
matrix elements in terms of the quark pole mass and the quarkonium
mass, and in \cite{0603186} the matrix elements are calculated based
on the potential model.
Differing from their methods, we get the production matrix elements
by using the experimentally observed charmonium decay rates. As
shown in\cite{9407339}, the difference between the color-singlet
production and decay matrix elements are of order $v^4$. So our
method should be valid at order $v^{2}$, and our estimates of the
production matrix elements should be good numerically, if the high
order QCD and $v^{2}$ corrections for the decays are small and the
experimental errors are not large.

\subsection{Numerical results and discussions}
At $\sqrt{s}=10.6$GeV, with $m_{c}=1.5$GeV and $\alpha_{s}=0.26$,
using the matrix elements in Eqs.[\ref{eqM1},\ref{eqM2}] determined
from charmonium decays with $v^2$ corrections but without QCD
radiative corrections, and making the phase space integral, the
leading order cross section of $e^{+}e^{-}\rightarrow
J/\psi+\eta_{c}$ (LO means for the short-distance part, since for
the long-distance matrix elements the $v^2$ corrections are already
included) is $3.07$ fb. The relativistic correction contributes
$0.798$ fb, which gives about $26\%$ enhancement and results in
$3.87$ fb for the cross section. The cross sections as functions of
$m_{c}$ are shown in Fig[2].  The lower line represents the LO
result in $v$, and the upper line represents the result with $v^2$
corrections. Since the long-distance matrix elements are
proportional to the squared quark mass, and the short distance
coefficients are found to be quite stable when $m_{c}$ changes, so
the cross section goes down as $m_{c}$ becomes smaller. However,
these results are obtained for fixed value of $\alpha_{s}=0.26$, and
if the running of $\alpha_{s}=\alpha_{s}(2m_c)$ is assumed, the
$m_c$ dependence of cross sections will be substantially changed,
making the cross section at $m_c$=1.4~GeV close to that at
$m_c$=1.6~GeV.

If using the matrix elements in Eqs.[\ref{eqM3},\ref{eqM4}]
determined from charmonium decays with both $v^2$ and $\alpha_{s}$
corrections, at order $v^{0}$  the cross section is $9.02$fb, and at
order $v^{2}$ the cross section is $11.26$fb for $\alpha_{s}=0.26$
and $m_{c}=1.5$Gev. When $m_{c}$ varies from $1.4$Gev to $1.6$Gev,
the cross section goes up from $9.18$fb to $13.43$fb.
\begin{figure}
\begin{center}
\includegraphics{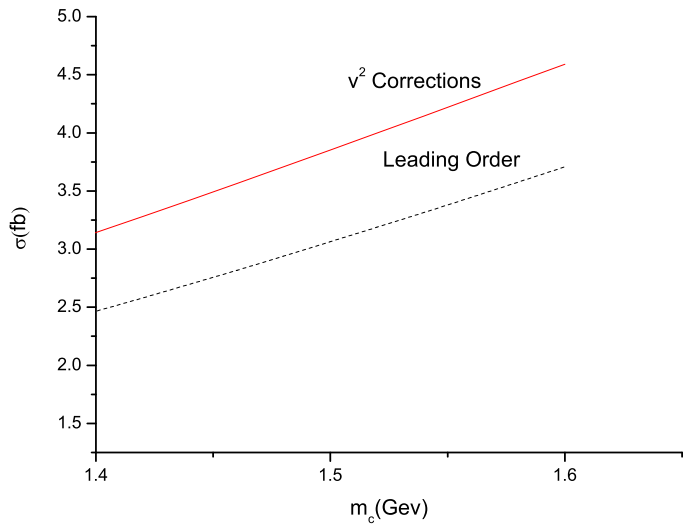}
\caption{$e^{+}e^{-}\rightarrow J/\psi+\eta_{c}$ cross sections with
relativistic corrections to long-distance matrix elements extracted
from charmonium decays (without NLO QCD radiative corrections). The
lower line represents the LO result in $v$, and the upper line
represents the result with $v^2$ corrections to the short-distance
coefficients. Here the coupling constant is fixed as
$\alpha_{s}=0.26$.}
\end{center}
\end{figure}

\begin{figure}
\begin{center}
\includegraphics{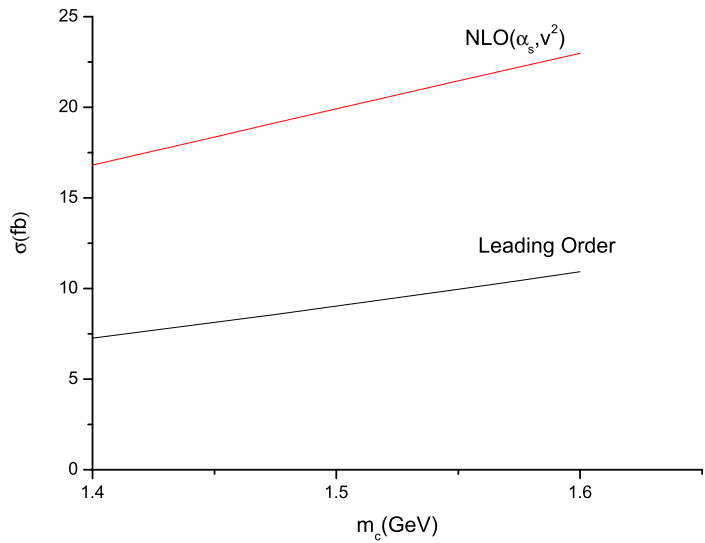}
\caption{$e^{+}e^{-}\rightarrow J/\psi+\eta_{c}$ cross sections with
relativistic corrections to long-distance matrix elements extracted
from charmonium decays (with NLO QCD radiative corrections). The
lower line represents the LO result in $v$, and the upper line
represents the result with $v^2$ corrections to the short-distance
coefficients. Here the coupling constant is fixed as
$\alpha_{s}=0.26$. Note that the QCD radiative corrections to the
short-distance coefficients (with K=1.8) are included for the upper
line but not the lower line. }
\end{center}
\end{figure}

The experiment result in Eq.[\ref{eq1}] is about an order of
magnitude larger than the leading order
result\cite{0211181,0211085,0305102}. The result at order $v^{2}$
shows that relativistic corrections can enhance both the short
distance coefficients and the long distance matrix elements. The
next-to-leading order $v^{2}$ coefficient
$(\overline{M}_{1J/\psi}+\overline{M}_{1\eta_{c}})m_{c}^{2}$ is
about 2.2 times larger than $\overline{M}$ for $m_{c}=1.5$Gev and
$\alpha_{s}=0.26$, and the matrix elements of
$\langle0|\mathcal{O}_{1}({}^{1}S_{0}^{\eta_c})|0\rangle$ and
$\langle0|\mathcal{O}_{1}({}^{3}S_{1}^{\psi})|0\rangle$ also become
about 1.17 times larger than that in the leading order calculation.
These make the cross section become 1.7 times larger after including
the relativistic effect. If we determine the matrix elements
including the QCD radiative correction, at $v^{2}$ order the
theoretical result is $11.26$fb for $\alpha_{s}=0.26$ and
$m_{c}=1.5$Gev. It shows that the relativistic corrections are
significant.

In the above discussions, we have not considered the QCD radiative
corrections to the short-distance coefficients as shown in
\cite{0506076}. If we further combine the NLO QCD corrections
\cite{0506076} with the $v^{2}$ corrections with a fixed value
$\alpha_{s}=0.26$, then the cross section will go from $16.8$fb to
$23.0$fb when $m_{c}$ varies from $1.4$GeV to $1.6$ GeV, as shown in
Fig[3]. The calculated cross sections of
$\sigma[e^{+}e^{-}\rightarrow J/\psi+\eta_{c}]$  are listed in Table
I, where $\sigma_{LO(\alpha_{s},v^2)}$ means the cross section at
leading order in both $\alpha_{s}$ and $v^{2}$;
$\sigma_{NLO(\alpha_{s})}$ means that obtained by using the short
distance part with $\alpha_{s}$ corrections \cite{0506076}, and the
long-distance matrix elements also with $\alpha_{s}$ corrections
extracted from both $J/\psi\rightarrow e^{+}e^{-}$ and
$\eta_{c}\rightarrow 2\gamma$ (not from $J/\psi\rightarrow
e^{+}e^{-}$ alone as was extracted in \cite{0506076}) with all at
leading order in $v^2$; $\sigma_{NLO(v^2 )}$ means that obtained
with $v^2$ corrections but without $\alpha_{s}$ corrections; and
$\sigma_{NLO(\alpha_{s},v^2)}$ means that obtained by combining both
$\alpha_{s}$ and $v^2$ corrections, where the matrix element are
taken from Eq.[\ref{eqM3}].

We see that with both QCD radiative corrections and relativistic
corrections, the discrepancy between experiment and theory for
$e^{+}e^{-}\rightarrow J/\psi+\eta_{c}$ could be largely resolved.
Our result is consistent with \cite{0603186}. In our approach the
long-distance matrix elements are extracted from experimental data
of $J/\psi$ and $\eta_c$ decays, and therefore are
model-independent. The main uncertainties may come from the higher
order corrections and the errors in the measurements. The
relativistic effects on the double charmonium production estimated
in our approach are milder than some results obtained by using the
light-cone methods.

\begin{table}
\caption{Experimental and calculated cross sections of
$\sigma[e^{+}e^{-}\rightarrow J/\psi+\eta_{c}]$ with $m_c=$1.5~GeV
and $\alpha_{s}$=0.26. See text for the definitions of
$\sigma_{LO(\alpha_{s},v^2)}$, $\sigma_{NLO(\alpha_{s})}$,
$\sigma_{NLO(v^2 )}$, and $\sigma_{NLO(\alpha_{s},v^2)}$.}


\small
\begin{tabular}{|c|c|c|c|c|c|}
     \hline
      \multicolumn{6}{|c|}{Experimental Result}\\
     \hline
     \multicolumn{3}{|c|}{$\sigma_{Belle}[e^{+}e^{-}\rightarrow J/\psi+\eta_{c}]\times
     \mathcal{B}^{\eta_{c}}[\geq2]$(fb)}&
     \multicolumn{3}{|c|}{$\sigma_{Babar}[e^{+}e^{-}\rightarrow J/\psi+\eta_{c}]\times
     \mathcal{B}^{\eta_{c}}[\geq2]$(fb)}\\
\hline
\multicolumn{3}{|c|}{$25.6\pm2.8\pm3.4$}& \multicolumn{3}{|c|}{$17.6\pm2.8\pm2.1$} \\
     \hline
     \multicolumn{6}{|c|}{Theoretical Result}\\
     \hline
$\langle0|\mathcal{O}_{1}({}^{1}S_{0}^{\eta_{c}})|0\rangle $ &
$\frac{\langle 0|\mathcal{O}_{1}({}^{3}S_{1}^{\psi})|0\rangle}{3}$ &
$\frac{\langle0|\mathcal{P}_{1}({}^{1}S_{0}^{\eta_{c}})|0\rangle}{m_{c}^{2}}$
&$\frac{\langle0|\mathcal{P}_{1}({}^{3}S_{1}^{\psi})|0\rangle}{3m_{c}^{2}}$
& $\alpha_{s}$
& $\sigma$ (fb)\\
\hline
0.243Gev$^{3}$ & 0.252Gev$^{3}$ & 0 & 0 & $\alpha_{s}=0.26$ & $\sigma_{LO(\alpha_{s},v^2)}=2.26$ \\
\hline
0.337Gev$^{3}$ & 0.450Gev$^{3}$ & 0 & 0 & $\alpha_{s}=0.26$ & $\sigma_{NLO(\alpha_{s})}=10.92$\\
\hline
0.286Gev$^{3}$ & 0.295Gev$^{3}$ & 0.0321Gev$^{3}$ & 0.0321Gev$^{3}$ & $\alpha_{s}=0.26$ &$\sigma_{NLO(v^2)}=3.87$\\
\hline
0.432Gev$^{3}$ & 0.573Gev$^{3}$ & 0.0514Gev$^{3}$ & 0.0514Gev$^{3}$& $\alpha_{s}=0.26$ & $\sigma_{NLO(\alpha_{s},v^2)}=20.04$\\
\hline
\end{tabular}
\end{table}

\section{Relativistic Corrections to $e^{+}e^{-}\rightarrow J/\psi+c\overline{c}$}
   The long distance matrix elements in process $e^{+}e^{-}\rightarrow
J/\psi+c\overline{c}$ are the same as in process
$e^{+}e^{-}\rightarrow J/\psi+\eta_{c}$, so we just use the result
above, and only give the detailed calculation for the short distance
part.
\subsection{Short distance part}
   There are four Feynman diagrams in the process
$e^{+}e^{-}\rightarrow (C\overline{C})_{{}^{3}S_{1}}+C\overline{C}$,
which are shown in Fig[4].
\begin{figure}
\begin{center}
\vspace{-3.0cm}
\includegraphics[width=14cm,height=14cm]{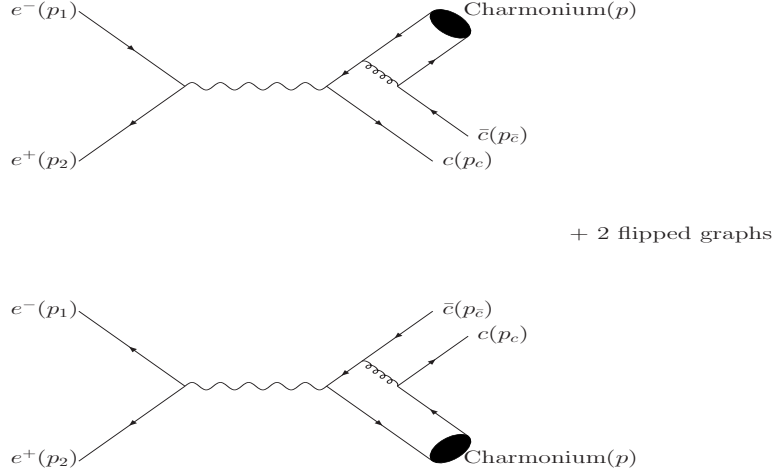}
\vspace{-4cm}
\end{center}
\caption{ Feynman diagrams for $e^+ +
e^-\rightarrow\gamma^*\rightarrow$ $J/\psi$ + $c\bar{c}$.}
\label{fey2}
\end{figure}
As in the last section the amplitude can be expanded in terms of the
quark relative momentum and reads
\begin{eqnarray}
M(e^{+}e^{-}\rightarrow
(C\overline{C})_{{}^{3}S_{1}}+C\overline{C})=
(\frac{m_{c}}{E})^{1/2}A(q_{\psi})=
\nonumber\\
(\frac{m_{c}}{E})^{1/2}(A(0)+q_{\psi}^{\alpha}\frac{\partial
A}{\partial q_{\psi}^{\alpha}}\Big{|}_{q_{\psi}=0}+
\frac{1}{2}q_{\psi}^{\alpha}q_{\psi}^{\beta}\frac{\partial^{2}
A}{\partial q_{\psi}^{\alpha}\partial
q_{\psi}^{\beta}}\Big{|}_{q_{\psi}=0}+\dots),
\end{eqnarray}
where
\begin{eqnarray}
A(q_{\psi})=\sum_{\lambda_{1}\lambda_{2}}\sum_{ij}
\langle\frac{1}{2},\lambda_{1},
\frac{1}{2},\lambda_{2}|1,S_{z}\rangle
\langle3,i;\overline{3},j|1\rangle A(e^{+}e^{-}\rightarrow
C_{\lambda1,i}(\frac{P}{2}+q_{\psi})\overline{C}_{\lambda2,j}(\frac{P}{2}-q_{\psi})+
C\overline{C}).
\end{eqnarray}
At leading order of $v^{2}$
\begin{eqnarray}
|M|^{2}=\frac{m_{c}}{E_{1}}A(0)A^{\ast}(0)+
\frac{1}{2}q_{\psi}^{\alpha}q_{\psi}^{\beta}A_{\alpha\beta}A^{\ast}(0)+
\frac{1}{2}q_{\psi}^{\alpha}q_{\psi}^{\beta}A_{\alpha\beta}^{\ast}A(0),
\end{eqnarray}
where $A_{\alpha\beta}=\frac{\partial^{2}A}{\partial
q^{\alpha}\partial q^{\beta}}$, and
$A_{\alpha\beta}^{\ast}=\frac{\partial^{2}A^{\ast}}{\partial
q^{\alpha}\partial q^{\beta}}$. Then
\begin{subequations}
\begin{equation}
\overline{N}_{0}=\frac{1}{2N_{c}m_{c}}(A(0)A^{\ast}(0))
\Big{|}_{\mathbf{q}_{J/\psi}^{2}=0},
\end{equation}
\begin{eqnarray}
\overline{N}_{1}=\frac{1}{2N_{c}m_{c}}(\frac{\partial
(\frac{m_{c}}{E}A(0)A^{\ast}(0))}{\partial
(\mathbf{q}_{\psi}^{2})}\Big{|}_{\mathbf{q}_{J/\psi}^{2}=0}
\nonumber\\+\frac{1}{6}(A_{\psi}^{\alpha\beta}\Pi_{\alpha\beta}(P_{J/\psi})A^{\ast}(0)+
A_{\psi}^{\ast\alpha\beta}\Pi_{\alpha\beta}(P_{J/\psi})A(0))
\Big{|}_{\mathbf{q}_{J/\psi}^{2}=0}).
\end{eqnarray}
\end{subequations}

The expressions of $A(0)A^{\ast}(0)$ and
$\Pi^{\alpha\beta}(A_{\alpha\beta}A^{\ast}(0)+A_{\alpha\beta}^{\ast}A(0))$
can be written in terms of some dimensionless variables
$z_{i}=2E_{i}/\sqrt{s},
\overrightarrow{q_{i}}=2\overrightarrow{p_{i}}/\sqrt{s},
x_{i}=\cos\theta_{i}$ and $\delta_{i}=2m_{i}/\sqrt{s}$. Here
$\sqrt{s}$ is the total energy in the center of mass frame,
$p_{3}^{\mu}$, $p_{4}^{\mu}$, $p_{5}^{\mu}$ are the four-momenta of
the final state $J/\psi$, charm quark and anticharm quark
respectively, $m_{i}^{2}=p_{i}^{2}$, and $\theta_{i}$ is the angle
between state $i$ and the electron. The scalar productions are
expressed as follows:
\begin{align}
&p_{1}.p_{3}=\frac{s}{4}(z_{3}-q_{3}x_{3});\;
p_{2}.p_{3}=\frac{s}{4}(z_{3}+q_{3}x_{3});\;
p_{4}.p_{5}=\frac{s}{8}(4-4z_{3}+\delta_{3}^2-\delta_{4}^{2}-\delta_{5}^{2});
\nonumber\\&p_{1}.p_{4}=\frac{s}{4}(z_{4}-(q_{-}x_{-}-q_{3}x_{3})/2);\;
p_{2}.p_{4}=\frac{s}{4}(z_{4}+(q_{-}x_{-}-q_{3}x_{3})/2);\;
\nonumber\\&p_{1}.p_{5}=\frac{s}{4}(z_{5}+(q_{-}x_{-}+q_{3}x_{3})/2);\;
p_{2}.p_{5}=\frac{s}{4}(z_{5}-(q_{-}x_{-}+q_{3}x_{3})/2);\;
\nonumber\\&p_{3}.p_{4}=\frac{s}{8}(4-\delta_{3}^{2}-4z_{5});\;
p_{3}.p_{5}=\frac{s}{8}(4-\delta_{3}^{2}-4z_{4});\;p_{1}.p_{2}=\frac{s}{2};
\end{align}
where $z_{-}=z_{4}-z_{5}$,
$q_{-}=|\overrightarrow{q_{4}}-\overrightarrow{q_{5}}|
=\sqrt{4-4z_{3}
+z_{-}^{2}+\delta_{3}^{2}-2\delta_{4}^{2}-2\delta_{5}^{2}}$,
$q_{3}=|\overrightarrow{q_{3}}|=\sqrt{z_{3}-\delta_{3}^2}$,
$x_{-}=\cos\theta_{-}$, and $\theta_{-}$ is the angle between
$\overrightarrow{q_{-}}$ and the electron.

\subsection{Numerical result}
In principle, the phase space calculation also contains relativistic
corrections (e.g., by inclusion of the binding energy in a meson).
But for simplicity, we calculate the phase space integration by
assuming $m_{J/\psi}=2m_{c}$ (the effect due to this simplification
is very small). Then the three-body final state phase space
expressed in terms of the variables defined above becomes
\begin{align}
d\Phi_{3}&=(2\pi)^{4} \;
\delta^{4}(p1+p2-p3-p4-p5)\prod_{i=3}^{5}\frac{d^{3}p_{i}}{2E_{i}}
\nonumber\\&=\frac{s}{32(2\pi)^4}\frac{dz_{3}dx_{3}dz_{-}dw}{\sqrt{(1-K^{2})(1-x_{3}^{2})-w^2}},
\end{align}
where
\begin{subequations}
\begin{equation}
K=\frac{z_{-}(2-z_{3})}{q_{3}q_{-}},
\end{equation}
\begin{equation}
w=x_{-}+Kx_{3}.
\end{equation}
\end{subequations}
The limits of those variables are
\begin{subequations}
\begin{equation}
\delta\leq z_{3}\leq 1,
\end{equation}
\begin{equation}
-1\leq x_{3} \leq1,
\end{equation}
\begin{equation}
-\sqrt{\frac{(z_{3}^{2}-\delta^{2})(4-4z_{3})}{4+\delta^{2}-4z_{3}}}
\leq
z_{-}\leq\sqrt{\frac{(z_{3}^{2}-\delta^{2})(4-4z_{3})}{4+\delta^{2}-4z_{3}}},
\end{equation}
\begin{equation}
-\sqrt{(1-K^{2})(1-x_{3}^{2})}\leq w
\leq\sqrt{(1-K^{2})(1-x_{3}^{2})},
\end{equation}
\end{subequations}
where $\delta=4m_{c}/\sqrt{s}$.

The expressions of $M_{0}$ and $M_{1}$ are lengthy, so we only give
the expression for the differential cross section. The differential
cross section of unpolarized $J/\psi$ production in association with
$c\bar c$ in $e^{+}e^{-}$ annihilation can be expressed as
\begin{equation}
\frac{d^{2}\sigma}{dE_{3}d\cos\theta_{3}}(e^{+}e^{-}\rightarrow
\gamma^{\ast}\rightarrow\psi+X)=S(E_{3})(1+\alpha\cos^{2}\theta_{3}),
\end{equation}
where $E_{3}$ is the energy of $J/\psi$, and $\theta_{3}$ is the
angle between $J/\psi$ and the electron. When expressing the above
equation by using the new parameters it becomes
\begin{equation}
\frac{d^{2}\sigma}{dz_{3}dx_{3}}(e^{+}e^{-}\rightarrow
\gamma^{\ast}\rightarrow\psi+X)=S(z_{3})(1+\alpha(z_{3}) x_{3}^{2}).
\end{equation}

The leading order result of $S(z_{3})$ and $\alpha(z_{3})$ are in
agreement with those in \cite{0301218,9606229}. We give the
next-to-leading order result of $S(z_{3})$ and $\alpha(z_{3})$ in
the Appendix \textbf{A}. Using the long-distance matrix elements in
Eq.[\ref{eqM1}] (without NLO $\alpha_s$ corrections in the
charmonium decay widths), the leading order cross section of
$e^{+}+e^{-}\rightarrow J/\psi+c\overline{c}$  with $m_{c}=1.5$ Gev,
$\sqrt{s}=10.6$ Gev, $\alpha_{s}=0.26$ is estimated to be $110$
fb,\footnote{This value is smaller than 148~fb given
in~\cite{0301218}, because here a smaller value of the wave function
squared at the origin is extracted from the $J/\psi$ data than that
used in \cite{0301218} from potential model calculations.} and the
next-to-leading order $v^{2}$ correction is only $0.42$ fb, which
gives about less than a half percent enhancement. As a result, the
relativistic correction to $e^{+}+e^{-}\rightarrow
J/\psi+c\overline{c}$ is found to be very small and negligible, in
contrast to the exclusive double charmonium production process
$e^{+}+e^{-}\rightarrow J/\psi+\eta_c$, in which the relativistic
correction is very significant. In fact, we find that in
$e^{+}+e^{-}\rightarrow J/\psi+c\overline{c}$ the short distance
part $\bar{N}_{1}m^{2}_{c}$ at order $v^{2}$ is much smaller than
the leading order term $\bar{N}_{0}$, i.e.,
$\frac{\bar{N}_{1}m^{2}_{c}}{\bar{N}_{0}}\ll1$. Here $\bar{N}_{1}$
in Eq.[25b] is the sum of two terms, and they both are small and
have different signs, so their sum becomes even tiny. So, the tiny
effect of relativistic corrections on the rate of
$e^{+}+e^{-}\rightarrow J/\psi+c\overline{c}$ is due to the
smallness of the short-distance coefficient correction, regardless
of the long-distance matrix elements.

When $\sqrt{s}$ is larger, the ratio of the correction at order
$v^2$ to the leading order contribution even changes sign from
positive to negative values, but its value is always small. This is
shown in Fig[5], where the values of the parameters are the same as
used above.

\begin{figure}
\begin{center}
\vspace{-2.0cm}
\includegraphics[width=14cm,height=8cm]{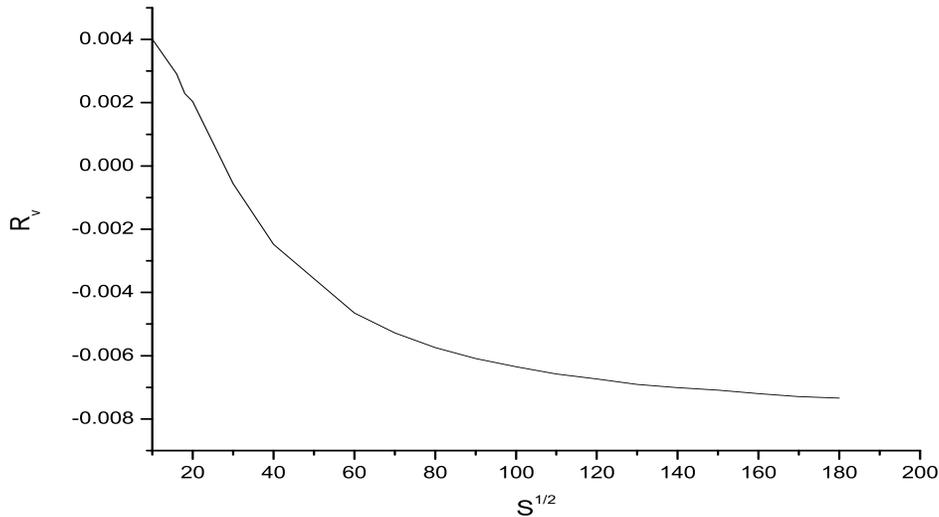}
\vspace{-0.1cm}
\end{center}
\caption{$R_{v}$ as a function of $\sqrt{s}$. Here $R_{v}$ is the
ratio of the correction at order $v^2$ to the leading order result
for the cross section of $e^{+}+e^{-}\rightarrow
J/\psi+c\overline{c}$.}
\end{figure}

If we use the enhanced matrix elements in Eq.[\ref{eqM3}], which are
obtained by including NLO $\alpha_s$ corrections in the charmonium
decay widths, for $m_{c}=1.5$Gev, $\sqrt{s}=10.6$Gev,
$\alpha_{s}=0.26$, the leading order result is about $214$fb, and
the relativistic correction at order $v^{2}$ is only $0.67$fb.

\subsection{Discussions}

Since for $e^{+}+e^{-}\rightarrow J/\psi+c\overline{c}$ the conflict
between experiment and theory at leading order in both $\alpha_{s}$
and $v^{2}$ is quite serious, a number of attempts have been made to
solve this problem. In \cite{0305084} the authors considered the
production of  $J/\psi+c\overline{c}$ through two photons, and found
that  the contribution of two photon process may be comparable to
that of one photon process when $\sqrt{s}$ is large, say, larger
than $\sqrt{s}=20$Gev, but at $\sqrt{s}=10.6$Gev, it is only 15\% of
the one photon contribution. Authors in \cite{0306062} considered
the process when the mass of the charm quark can be neglected. In
\cite{0303145} the result in factorization method  was compared with
the duality method. Other suggestions can also be found in e.g.
\cite{Kaidalov:2003wp,Kang:2004zj}. Despite of these efforts, a
satisfactory resolution is still needed. Recently, the authors of
ref.\cite{0611086} have calculated the next-to-leading order QCD
corrections to the direct $J/\psi+c\bar{c}+X$ production in
$e^{+}+e^{-}$ annihilation at $\sqrt{s}=10.6$GeV, and with
$m_{c}=1.5$GeV and $\alpha_{s}=0.26$ their result is
\begin{equation}
\sigma_{direct}(e^{+}e^{-}\rightarrow J/\psi+c\overline{c})\approx
0.33\textrm{pb},
\end{equation}
It enhances the leading order result by a factor of $1.8$. Further
including the contributions from $\psi(2S)$ and other states, the
cross section of prompt $J/\psi+c\overline{c}$ production becomes
\begin{equation}
\sigma_{prompt}(e^{+}e^{-}\rightarrow J/\psi+c\overline{c}+X)\approx
0.51\textrm{pb}.
\end{equation}
It is about $60\%$ of the Belle value in Eq.[\ref{eq2}]. This result
shows that the QCD radiative corrections to $e^{+}+e^{-}\rightarrow
J/\psi+c\overline{c}$ is essential.

Here, when we further consider the relativistic corrections at order
$v^{2}$, we find that in contrast to the $J/\psi+\eta_{c}$
production, the relativistic corrections to $J/\psi+c\overline{c}$
production are very small. The relativistic corrections can only
improve the leading order result by enlarging the long distance
matrix elements, and have little effect on the short distance part.
As a result, the relativistic corrections to $e^{+}+e^{-}\rightarrow
J/\psi+c\overline{c}$ are tiny and negligible.

The physical reason for the difference between exclusive and
inclusive processes largely lies in the fact that in the exclusive
process $e^{+}e^{-}\rightarrow J/\psi+\eta_{c}$ the virtuality of
the gluon that converts into a charm quark pair takes its maximum
value $\frac{s}{4}$ in the nonrelativistic limit (with zero relative
momentum between quarks in charmonium), and introducing the relative
momentum can substantially reduce the gluon virtuality, and hence
enhance the short-distance coefficient and the cross section;
whereas for the inclusive process $e^{+}+e^{-}\rightarrow
J/\psi+c\overline{c}$, the charm quark fragmentation $c\to J/\psi+c$
($\bar c\to J/\psi+\bar c$) is significant, in which the quark
relative momentum has little effect on the virtuality of the gluon.
Therefore, our result for the next-to-leading $v^2$ corrections may
indicate that the relativistic correction is not a good direction to
solve the problem of discrepancy  between theory and experiment in
$J/\psi+c\bar{c}$ production. Further studies for QCD radiative
corrections and other possible new mechanisms related to the double
charm production could be more useful.

\section{Summary}
In this paper, we have studied the relativistic corrections at order
$v^{2}$ to the double-charm production processes
$e^{+}e^{-}\rightarrow J/\psi+\eta_{c}$ and $e^{+}e^{-}\rightarrow
J/\psi+c\overline{c}$ at B factories in the framework of
non-relativistic quantum chromodynamics. The short-distance parts of
production cross sections are calculated perturbatively, while the
long-distance matrix elements are extracted from experimental data
for $J/\psi$ and $\eta_c$ decays up to errors of order $v^4$, and
therefore are model-independent. The main uncertainties may come
from the higher order corrections and the errors in the
measurements. Our results show that the relativistic correction to
the exclusive process $e^{+}e^{-}\rightarrow J/\psi+\eta_{c}$ is
significant, which, when combined together with the next-to-leading
order $\alpha_{s}$ corrections, could resolve the large discrepancy
between theory and experiment. It can be clearly seen from TABLE I
that for the cross section the relativistic correction alone gives
an enhancement factor of 1.7 while the combination of relativistic
correction with QCD radiative correction results in a much larger
enhancement factor of 9. This conclusion is consistent with
\cite{0603186}, in which, however, the long-distance matrix elements
are estimated by using potential model calculations. The
relativistic effects on the $J/\psi\eta_c$ production estimated in
our approach are milder than some results obtained by using the
light-cone methods. On the other hand, for the inclusive process
$e^{+}e^{-}\rightarrow J/\psi+c\overline{c}$ we find that the
relativistic correction is tiny and negligible, and therefore not
helpful in resolving the discrepancy between theory and experiment.
The physical reason for the above difference between exclusive and
inclusive processes largely lies in the fact that in the exclusive
process the relative momentum between quarks in charmonium
substantially reduces the virtuality of the gluon that converts into
a charm quark pair, but this is not the case for the inclusive
process, in which the charm quark fragmentation $c\to J/\psi+c$
($\bar c\to J/\psi+\bar c$) is significant, and QCD radiative
corrections can be more essential. Further studies are needed to
understand the large ratio of the cross section of
$e^{+}e^{-}\rightarrow J/\psi+c\overline{c}$ to the cross section of
$e^{+}e^{-}\rightarrow J/\psi+anything$, which is measured to be in
the range of 0.6-0.8 by Belle.

\begin{acknowledgments}
We thank G.T. Bodwin for useful comments. This work was supported in
part by the National Natural Science Foundation of China (No
10421503, No 10675003), the Key Grant Project of Chinese Ministry of
Education (No 305001), and the Research Found for Doctorial Program
of Higher Education of China.
\end{acknowledgments}

\appendix

\section{}

The formulas of $S(z_{3})$ and $\alpha_(z_{3})$ are given as
follows:
\begin{align}
S(z_{3})&=\frac{2e_{c}^{2}\pi\alpha^2\alpha_{s}^2}
{3^{5}s^{3}\delta^4z_{3}^{4}\sqrt{1-z_{3}}(2-z_{3})^6(z_{3}^2-\delta^2)}
\frac{\langle0|\mathcal{P}_{1}({}^{3}S_{1}^{\psi})|0\rangle}{m_{c}}\times
\nonumber\\&\ \Big(-(98304\,{z_{3}}^5 + 98304\,{z_{3}}^6 +
696320\,{z_{3}}^7 - 6963200\,{z_{3}}^8 + 14895104\,{z_{3}}^9 -
13801472\,{z_{3}}^{10} + \nonumber\\&6104576\,{z_{3}}^{11} -
1221632\,{z_{3}}^{12} + 93696\,{z_{3}}^{13} -
737280\,{z_{3}}^3\,{\delta }^2 + 3047424\,{z_{3}}^4\,{\delta }^2 -
6987776\,{z_{3}}^5\,{\delta }^2 +
\nonumber\\&15003648\,{z_{3}}^6\,{\delta }^2 -
22528000\,{z_{3}}^7\,{\delta }^2 + 17085440\,{z_{3}}^8\,{\delta }^2-
4745728\,{z_{3}}^9\,{\delta }^2 - 411904\,{z_{3}}^{10}\,{\delta}^2 +
\nonumber\\&299584\,{z_{3}}^{11}\,{\delta }^2 -
26176\,{z_{3}}^{12}\,{\delta }^2+ 737280\,z_{3}\,{\delta }^4 -
4030464\,{z_{3}}^2\,{\delta}^4 + 7950336\,{z_{3}}^3\,{\delta }^4 -
\nonumber\\&7868416\,{z_{3}}^4\,{\delta }^4 +
3901440\,{z_{3}}^5\,{\delta }^4 + 2526208\,{z_{3}}^6\,{\delta }^4 -
6324224\,{z_{3}}^7\,{\delta }^4 + 3657984\,{z_{3}}^8\,{\delta }^4 -
\nonumber\\&540096\,{z_{3}}^9\,{\delta }^4 -
42816\,{z_{3}}^{10}\,{\delta }^4 + 3632\,{z_{3}}^{11}\,{\delta }^4 +
442368\,{\delta }^6 - 1124352\,z_{3}\,{\delta }^6 +
387072\,{z_{3}}^2\,{\delta }^6 +
\nonumber\\&948224\,{z_{3}}^3\,{\delta }^6 -
1466368\,{z_{3}}^4\,{\delta }^6 + 1291008\,{z_{3}}^5\,{\delta }^6 -
68864\,{z_{3}}^6\,{\delta }^6 - 493440\,{z_{3}}^7\,{\delta }^6 +
99136\,{z_{3}}^8\,{\delta }^6 +
\nonumber\\&34552\,{z_{3}}^9\,{\delta }^6 -
664\,{z_{3}}^{10}\,{\delta }^6 + 211968\,{\delta }^8 -
489984\,z_{3}\,{\delta }^8 + 357888\,{z_{3}}^2\,{\delta }^8 -
14336\,{z_{3}}^3\,{\delta }^8 -
\nonumber\\&365056\,{z_{3}}^4\,{\delta }^8 +
357696\,{z_{3}}^5\,{\delta }^8 - 42464\,{z_{3}}^6\,{\delta }^8 -
23104\,{z_{3}}^7\,{\delta }^8 - 7356\,{z_{3}}^8\,{\delta }^8 -
138\,{z_{3}}^9\,{\delta }^8 + \nonumber\\& 39168\,{\delta }^{10} -
93696\,z_{3}\,{\delta }^{10} + 95232\,{z_{3}}^2\,{\delta }^{10} -
20224\,{z_{3}}^3\,{\delta }^{10}- 52576\,{z_{3}}^4\,{\delta }^{10} +
23392\,{z_{3}}^5\,{\delta}^{10}+
\nonumber\\&3904\,{z_{3}}^6\,{\delta }^{10} +
432\,{z_{3}}^7\,{\delta }^{10} + 45\,{z_{3}}^8\,{\delta }^{10} +
3456\,{\delta }^{12} - 9216\,z_{3}\,{\delta }^{12} +
8352\,{z_{3}}^2\,{\delta }^{12} +
\nonumber\\&384\,{z_{3}}^3\,{\delta }^{12} -
2520\,{z_{3}}^4\,{\delta }^{12} + 192\,{z_{3}}^5\,{\delta }^{12} -
90\,{z_{3}}^6\,{\delta }^{12})
\frac{4z_{3}(1-z_{3})}{3(2-z_{3})^2}\sqrt{\frac{z_3^2-\delta^2}{\delta^2-4z_{3}+4}}
\nonumber\\&-192(256 - 512\,z_{3} + 320\,{z_{3}}^2 - 64\,{z_{3}}^3 +
16\,z_{3}^4 - 16\,z_{3}^5 + 4\,z_{3}^6 - 128\,{\delta }^2 +
64\,z_{3}\,{\delta }^2 +  \nonumber\\& 240\,z_{3}^2\,{\delta }^2
-224\,z_{3}^3\,{\delta }^2 + 20\,z_{3}^4\,{\delta }^2 +
16\,z_{3}^5\,{\delta }^2 + 48\,{\delta }^4 - 96\,z_{3}\,{\delta }^4
+ 24\,z_{3}^2\,{\delta }^4 + 24\,z_{3}^3\,{\delta }^4 +
3\,z_{3}^4\,{\delta }^4)\nonumber\\&\delta^{4}z_{3}^{4}(1-z_{3})
\arctan\frac{z_3^2-\delta^2}{\delta^2-4z_{3}+4} \nonumber\\
(&8192\,{z_{3}}^6 - 102400\,z_{3}^7 + 262144\,z_{3}^8 -
282624\,z_{3}^9 + 146944\,z_{3}^{10} - 35584\,z_{3}^{11} +
3072\,z_{3}^{12} + \nonumber\\&69632\,z_{3}^3\,{\delta }^2 -
299008\,z_{3}^4\,{\delta }^2 + 542720\,z_{3}^5\,{\delta }^2 -
564224\,z_{3}^6\,{\delta }^2 + 331264\,z_{3}^7\,{\delta }^2 -
87808\,z_{3}^8\,{\delta }^2 + \nonumber\\& 1664\,z_{3}^9\,{\delta
}^2 + 1472\,z_{3}^{10}\,{\delta }^2 + 336\,z_{3}^{11}\,{\delta }^2 -
61440\,z_{3}\,{\delta }^4 + 299008\,z_{3}^2\,{\delta }^4 -
486400\,z_{3}^3\,{\delta }^4 + \nonumber\\&388096\,z_{3}^4\,{\delta
}^4 - 180480\,z_{3}^5\,{\delta }^4 +  65280\,z_{3}^6\,{\delta }^4 -
19008\,z_{3}^7\,{\delta}^4 - 2880\,z_{3}^8\,{\delta }^4 +
3264\,z_{3}^9\,{\delta }^4 - \nonumber\\&288\,z_{3}^{10}\,{\delta
}^4 - 36864\,{\delta }^6 + 72192\,z_{3}\,{\delta }^6 -
16896\,z_{3}^2\,{\delta }^6 - 64256\,z_{3}^3\,{\delta}^6 +
65664\,z_{3}^4\,{\delta}^6 - \nonumber\\&15360\,z_{3}^5\,{\delta}^6-
6240\,z_{3}^6\,{\delta }^6 + 304\,z_{3}^7\,{\delta }^6 +
2216\,z_{3}^8\,{\delta }^6 + 6\,z_{3}^9\,{\delta }^6 -
8448\,{\delta}^8 +  14336\,z_{3}\,{\delta}^8 -
\nonumber\\&6144\,z_{3}^2\,{\delta }^8 - 4864\,z_{3}^3\,{\delta }^8
+ 1440\,z_{3}^4\,{\delta }^8 + 2624\,z_{3}^5\,{\delta }^8 -
1056\,z_{3}^6\,{\delta }^8 + 15\,z_{3}^8\,{\delta }^8 - \nonumber\\&
1152\,{\delta }^{10} + 3072\,z_{3}\,{\delta }^{10} -
2592\,z_{3}^2\,{\delta }^{10} + 1344\,z_{3}^3\,{\delta }^{10} -
696\,z_{3}^4\,{\delta }^{10} + 240\,z_{3}^5\,{\delta }^{10} -
30\,z_{3}^6\,{\delta }^{10})\nonumber\\& \delta^2\sqrt{1-z_{3}}
\ln\frac{z_{3}\sqrt{\delta^2-4z_{3}+4}+2\sqrt{(1-z_{3})(z_3^2-\delta^2)}}
{z_{3}\sqrt{\delta^2-4z_{3}+4}-2\sqrt{(1-z_{3})(z_3^2-\delta^2)}}\Big),
\end{align}
\begin{align}
\alpha(&z_{3})S(z_{3})=\frac{2e_{c}^{2}\pi\alpha^2\alpha_{s}^2}
{3^{5}s^{3}\delta^4z_{3}^{4}\sqrt{1-z_{3}}(2-z_{3})^6(z_{3}^2-\delta^2)}
\frac{\langle0|\mathcal{P}_{1}({}^{3}S_{1}^{\psi})|0\rangle}{m_{c}}\times
\nonumber\\\Big(&-(98304\,{z_{3}}^7 + 98304\,z_{3}^8 +
696320\,z_{3}^9 - 6963200\,z_{3}^{10} + 14895104\,z_{3}^{11} -
\nonumber\\& 13801472\,z_{3}^{12} + 6104576\,z_{3}^{13} -
1221632\,z_{3}^{14} + 93696\,z_{3}^{15} -
835584\,z_{3}^5\,{\delta}^2 + 3342336\,z_{3}^6\,{\delta }^2 -
\nonumber\\& 11059200\,z_{3}^7\,{\delta }^2 +
31518720\,z_{3}^8\,{\delta }^2 - 49162240\,z_{3}^9\,{\delta }^2 +
37194752\,z_{3}^{10}\,{\delta }^2 -
11941888\,z_{3}^{11}\,{\delta}^2+ \nonumber\\&
816896\,z_{3}^{12}\,{\delta }^2 + 153152\,z_{3}^{13}\,{\delta }^2 -
26176\,z_{3}^{14}\,{\delta }^2 - 1867776\,z_{3}^3\,{\delta }^4 +
6881280\,z_{3}^4\,{\delta }^4 - \nonumber\\&
4657152\,z_{3}^5\,{\delta }^4 - 19660800\,z_{3}^6\,{\delta }^4 +
43751424\,z_{3}^7\,{\delta }^4 - 31881216\,z_{3}^8\,{\delta }^4 +
6889984\,z_{3}^9\,{\delta }^4 + \nonumber\\&
366080\,z_{3}^{10}\,{\delta }^4 + 119808\,z_{3}^{11}\,{\delta }^4 +
31744\,z_{3}^{12}\,{\delta }^4 - 976\,z_{3}^{13}\,{\delta }^4 +
245760\,z_{3}\,{\delta }^6 - 1622016\,z_{3}^2\,{\delta}^6
+\nonumber\\& 3229696\,z_{3}^3\,{\delta }^6 -
1411072\,z_{3}^4\,{\delta }^6 + 2289664\,z_{3}^5\,{\delta }^6 -
12958720\,z_{3}^6\,{\delta }^6 + 16642816\,z_{3}^7\,{\delta }^6
-\nonumber\\& 7645696\,z_{3}^8\,{\delta }^6 +
1814848\,z_{3}^9\,{\delta }^6 - 524672\,z_{3}^{10}\,{\delta }^6 +
16520\,z_{3}^{11}\,{\delta }^6 - 1240\,z_{3}^{12}\,{\delta }^6 +
147456\,{\delta }^8 - \nonumber\\&522240\,z_{3}\,{\delta }^8 +
967680\,z_{3}^2\,{\delta }^8 - 2955776\,z_{3}^3\,{\delta }^8 +
6496768\,z_{3}^4\,{\delta }^8 -
5705984\,z_{3}^5\,{\delta}^8+1570816\,z_{3}^6\,{\delta }^8 -
\nonumber\\&345536\,z_{3}^7\,{\delta }^8 + 177184\,z_{3}^8\,{\delta
}^8 + 68040\,z_{3}^9\,{\delta}^8 + 3420\,z_{3}^{10}\,{\delta }^8 +
294\,z_{3}^{11}\,{\delta }^8 +  46080\,{\delta }^{10} +
10752\,z_{3}\,{\delta }^{10} -\nonumber\\&
254208\,z_{3}^2\,{\delta}^{10} + 92672\,z_{3}^3\,{\delta }^{10} +
144640\,z_{3}^4\,{\delta }^{10} + 86976\,z_{3}^5\,{\delta }^{10} +
9920\,z_{3}^6\,{\delta }^{10} - 81632\,z_{3}^7\,{\delta}^{10} -
\nonumber\\&996\,z_{3}^8\,{\delta }^{10} - 2070\,z_{3}^9\,{\delta
}^{10} + 45\,z_{3}^{10}\,{\delta }^{10} + 16128\,{\delta }^{12} -
18432\,z_{3}\,{\delta }^{12} - 15744\,z_{3}^2\,{\delta }^{12} +
65408\,z_{3}^3\,{\delta }^{12} -
\nonumber\\&116416\,z_{3}^4\,{\delta }^{12} +
61312\,z_{3}^5\,{\delta }^{12} - 4136\,z_{3}^6\,{\delta }^{12} +
1800\,z_{3}^7\,{\delta }^{12} + 45\,z_{3}^8\,{\delta }^{12} +
3456\,{\delta }^{14} - 9216\,z_{3}\,{\delta }^{14} +
\nonumber\\&8352\,z_{3}^2\,{\delta }^{14} + 384\,z_{3}^3\,{\delta
}^{14} - 2520\,z_{3}^4\,{\delta }^{14} + 192\,z_{3}^5\,{\delta
}^{14} - 90\,z_{3}^6\,{\delta}^{14})
\frac{4z_{3}(1-z_{3})}{3(2-z_{3})^2}\sqrt{\frac{z_3^2-\delta^2}{\delta^2-4z_{3}+4}}
\nonumber\\&+192(256 - 512\,z_{3} + 320\,z_{3}^2 - 64\,z_{3}^3 +
16\,z_{3}^4 - 16\,z_{3}^5 + 4\,z_{3}^6 - 128\,{\delta }^2 +
64\,z_{3}\,{\delta }^2 + 240\,z_{3}^2\,{\delta }^2 -
\nonumber\\&224\,z_{3}^3\,{\delta }^2 + 20\,z_{3}^4\,{\delta }^2 +
16\,z_{3}^5\,{\delta }^2 + 48\,{\delta }^4 - 96\,z_{3}\,{\delta }^4
+ 24\,z_{3}^2\,{\delta }^4 + 24\,z_{3}^3\,{\delta }^4 +
3\,z_{3}^4\,{\delta }^4) \nonumber\\ &\delta^{4}z_{3}^{4}(1-z_{3})
\arctan\frac{z_3^2-\delta^2}{\delta^2-4z_{3}+4}+(32768\,z_{3}^7 -
90112\,z_{3}^8 + 20480\,z_{3}^9 + 180224\,z_{3}^{10} -\nonumber\\&
251904\,z_{3}^{11} + 140800\,z_{3}^{12} - 35072\,z_{3}^{13} +
3072\,z_{3}^{14} -  258048\,z_{3}^5\,{\delta }^2 +
806912\,z_{3}^6\,{\delta }^2 - 845824\,z_{3}^7\,{\delta }^2 +
\nonumber\\&87040\,z_{3}^8\,{\delta }^2 +
484864\,z_{3}^9\,{\delta}^2 - 382208\,z_{3}^{10}\,{\delta }^2 +
116096\,z_{3}^{11}\,{\delta}^2 - 12608\,z_{3}^{12}\,{\delta }^2 -
48\,z_{3}^{13}\,{\delta }^2 + \nonumber\\&49152\,z_{3}^3\,{\delta}^4
- 286720\,z_{3}^4\,{\delta}^4 + 642048\,z_{3}^5\,{\delta }^4 -
845824\,z_{3}^6\,{\delta }^4 + 871680\,z_{3}^7\,{\delta }^4 -
684544\,z_{3}^8\,{\delta }^4 +
\nonumber\\&341824\,z_{3}^9\,{\delta}^4-
90112\,z_{3}^{10}\,{\delta}^4 + 9712\,z_{3}^{11}\,{\delta }^4 -
288\,z_{3}^{12}\,{\delta }^4 - 20480\,z_{3}\,{\delta }^6 +
122880\,z_{3}^2\,{\delta }^6 -
\nonumber\\&289280\,z_{3}^3\,{\delta}^6 +
453120\,z_{3}^4\,{\delta}^6 - 582144\,z_{3}^5\,{\delta }^6 +
489856\,z_{3}^6\,{\delta }^6 - 242112\,z_{3}^7\,{\delta }^6 +
64864\,z_{3}^8\,{\delta }^6 - \nonumber\\& 6352\,z_{3}^9\,{\delta
}^6 - 1336\,z_{3}^{10}\,{\delta }^6 - 138\,z_{3}^{11}\,{\delta }^6 -
12288\,{\delta }^8 + 36352\,z_{3}\,{\delta }^8 -
70400\,z_{3}^2\,{\delta }^8 +  136448\,z_{3}^3\,{\delta }^8
-\nonumber\\& 160640\,z_{3}^4\,{\delta }^8 +
127616\,z_{3}^5\,{\delta }^8 - 62400\,z_{3}^6\,{\delta }^8 +
12272\,z_{3}^7\,{\delta}^8+ 1256\,z_{3}^8\,{\delta }^8 -
6\,z_{3}^9\,{\delta }^8 + 15\,z_{3}^{10}\,{\delta }^8 -
\nonumber\\&768\,{\delta }^{10} - 10752\,z_{3}\,{\delta }^{10} +
29568\,z_{3}^2\,{\delta }^{10} - 36480\,z_{3}^3\,{\delta}^{10} +
18368\,z_{3}^4\,{\delta }^{10} - 2144\,z_{3}^5\,{\delta }^{10} -
472\,z_{3}^6\,{\delta }^{10} - \nonumber\\&24\,z_{3}^7\,{\delta
}^{10} + 15\,z_{3}^8\,{\delta}^{10} - 1152\,{\delta }^{12} +
3072\,z_{3}\,{\delta }^{12} - 2592\,z_{3}^2\,{\delta }^{12} +
1344\,z_{3}^3\,{\delta }^{12} -696\,z_{3}^4\,{\delta }^{12}
+\nonumber\\& 240\,z_{3}^5\,{\delta }^{12} - 30\,z_{3}^6\,{\delta
}^{12})\delta^2\sqrt{1-z_{3}}\ln\frac{z_{3}\sqrt{\delta^2-4z_{3}+4}+2\sqrt{(1-z_{3})(z_3^2-\delta^2)}}
{z_{3}\sqrt{\delta^2-4z_{3}+4}-2\sqrt{(1-z_{3})(z_3^2-\delta^2)}}\Big).
\end{align}

\end{document}